# Structural diverseness of neurons between brain areas and between cases

Ryuta Mizutani[1]*, Rino Saiga[1], Yoshiro Yamamoto[2], Masayuki Uesugi[3], Akihisa Takeuchi[3], Kentaro Uesugi[3], Yasuko Terada[3], Yoshio Suzuki[4], Vincent De Andrade[5], Francesco De Carlo[5], Susumu Takekoshi[6], Chie Inomoto[6], Naoya Nakamura[6], Youta Torii[7], Itaru Kushima[7,8], Shuji Iritani[7,9], Norio Ozaki[7], Kenichi Oshima[9,10], Masanari Itokawa[9,10], and Makoto Arai[10]

[1]Department of Applied Biochemistry, Tokai University, Hiratsuka, Kanagawa 259-1292, Japan
[2]Department of Mathematics, Tokai University, Hiratsuka, Kanagawa 259-1292, Japan
[3]Japan Synchrotron Radiation Research Institute (JASRI/SPring-8), Sayo, Hyogo 679-5198, Japan
[4]Photon Factory, High Energy Accelerator Research Organization KEK, Tsukuba, Ibaraki 305-0801, Japan
[5]Advanced Photon Source, Argonne National Laboratory, Lemont, IL 60439, USA
[6]Tokai University School of Medicine, Isehara, Kanagawa 259-1193, Japan
[7]Department of Psychiatry, Nagoya University Graduate School of Medicine, Nagoya, Aichi 466-8550, Japan
[8]Medical Genomics Center, Nagoya University Hospital, Nagoya 466-8550, Aichi, Japan
[9]Tokyo Metropolitan Matsuzawa Hospital, Setagaya, Tokyo 156-0057, Japan
[10]Tokyo Metropolitan Institute of Medical Science, Setagaya, Tokyo 156-8506, Japan

*Correspondence address:
Department of Applied Biochemistry, Tokai University, Hiratsuka, Kanagawa 259-1292, Japan
Phone: +81-463-58-1211; Fax: +81-463-50-2426
E-mail: mizutanilaboratory@gmail.com

**Abstract**

The cerebral cortex is composed of multiple cortical areas that exert a wide variety of brain functions. Although human brain neurons are genetically and areally mosaic, the three-dimensional structural differences between neurons in different brain areas or between the neurons of different individuals have not been delineated. Here, we report a nanometer-scale geometric analysis of brain tissues of the superior temporal gyrus of 4 schizophrenia and 4 control cases by using synchrotron radiation nanotomography. The results of the analysis and a comparison with results for the anterior cingulate cortex indicated that 1) neuron structures are dissimilar between brain areas and that 2) the dissimilarity varies from case to case. The structural diverseness was mainly observed in terms of the neurite curvature that inversely correlates with the diameters of the neurites and spines. The analysis also revealed the geometric differences between the neurons of the schizophrenia and control cases, suggesting that neuron structure is associated with brain function. The area dependency of the neuron structure and its diverseness between individuals should represent the individuality of brain functions.

The human cerebral cortex is composed of multiple cortical areas that exert a wide variety of brain functions, such as cognitive functions by the prefrontal cortex and auditory functions by the temporal lobe. Brodmann divided the human cerebral cortex into 52 cortical areas distinguished by their cytoarchitectures and by taking into account functional localization (Brodmann, 1909; Amunts & Zilles, 2015). These structurally and functionally different brain areas supposedly operate as modules (Sporns & Betzel, 2016) that are organized differently between individuals (Passingham et al., 2002; Finn et al., 2015; Mars et al., 2018). Such large-scale differences between individuals should originate from microscopic differences in neurons that constitute our brain. Although it has been reported that human brain neurons are not monoclonal but rather genetically and areally mosaic due to somatic mutations during developmental stages (Lodato et al., 2015; McConnell et al., 2017), the three-dimensional structural differences between neurons in different brain areas or between neurons of different individuals have not been delineated.

We previously reported a three-dimensional structural study of neurons of the anterior cingulate cortex of schizophrenia cases and controls (Mizutani et al., 2019). The geometric analysis of the neurons indicated that neurite curvature significantly differs between individuals and the difference become extraordinary in schizophrenia. It has been reported that the anterior cingulate cortex is responsible for emotional and cognitive functions (Botvinick et al., 1999; Bush et al., 2000) and is associated with mental disorders including schizophrenia (Bouras et al., 2001; Fornito et al., 2009). Therefore, the structural differences of neurons in the anterior cingulate cortex should lead to differences in the microcircuits that are relevant to cognitive functions and mental disorders. A further analysis of neuron structures in other brain areas will reveal how our neuron structures differ between brain areas or between cases.

In this study, we analyzed nanometer-scale three-dimensional structures of neurons of Brodmann area 22 (BA22) of the superior temporal gyrus (Fig. 1) of 4 schizophrenia cases (hereafter called S1–S4) and age/gender-matched 4 control cases (N1–N4) by using synchrotron radiation nanotomography (Supplementary Tables S1–S3; Takeuchi et al., 2002; De Andrade et al., 2016; Suzuki et al., 2016; Supplementary materials will be provided separately). The cases used in this study are the same as those of our previous report regarding the neuron structure of the BA24 area of the anterior cingulate cortex (Mizutani et al., 2019). This allowed us to compare neuron structures between the brain areas within each individual. The BA22 and BA24 tissues used in our studies were taken from the left hemispheres of biopsied brains and stained with Golgi impregnation. A total of 34 three-dimensional image datasets of layer V of the BA22 cortex were blinded by coding dataset names and subjected to a computerized procedure (Mizutani et al., 2019) to build Cartesian coordinate models of tissue structures (Fig. 1; Supplementary Video S1; Supplementary Data S1–S3). After the model-building procedures for all datasets were completed, the blinded datasets were re-assigned to the cases and analyzed by calculating geometric parameters including curvature and torsion of neurites, which represent the reciprocal of the radius of three-dimensional neurite curve and the deviation of the curve from a plane, respectively.

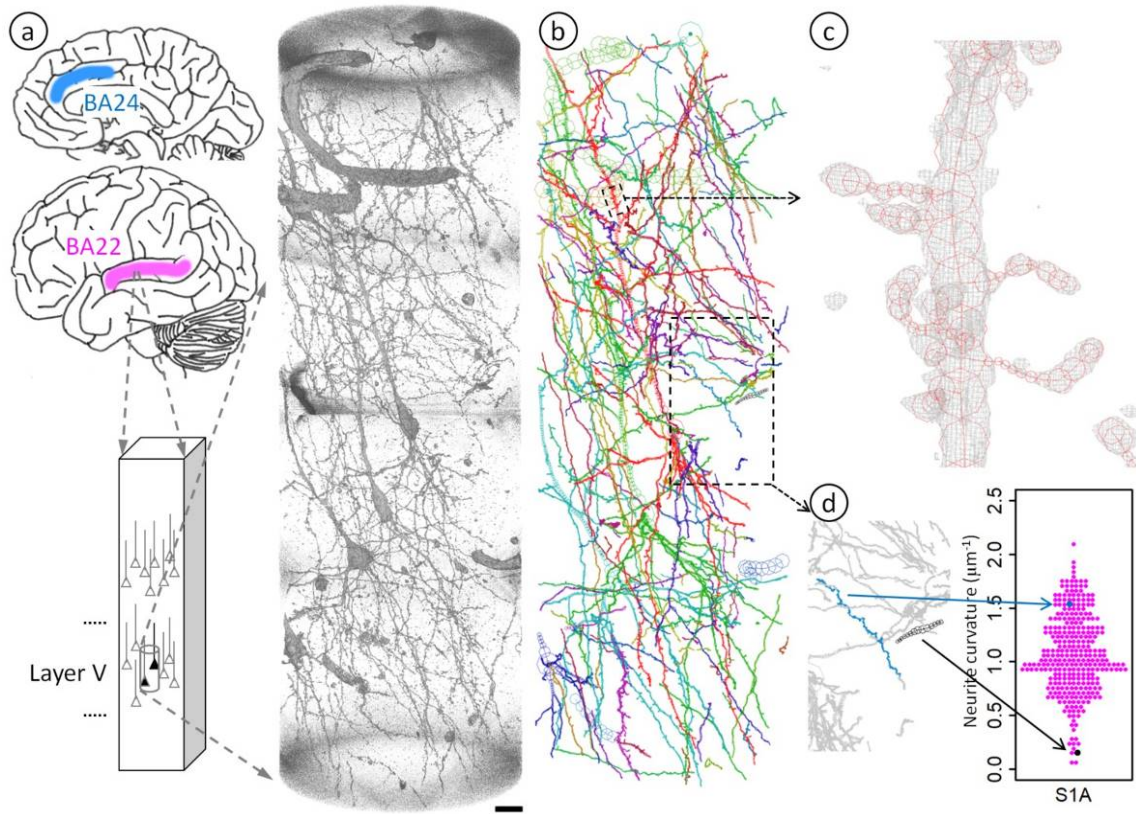

**Figure 1.** (**a**) Brain tissues of Brodmann area 22 (BA22, magenta) of the temporal lobe were stained with Golgi impregnation and subjected to synchrotron radiation nanotomography to visualize three-dimensional tissue structures in layer V. The obtained datasets along with those of the BA24 area (blue) of the anterior cingulate cortex were used for the geometric analysis. The rendering shows a three-dimensional image of the S1A dataset taken from BA22 tissue of the schizophrenia S1 case. The pial surface is toward the top. Voxel values of 160–800 were rendered with the scatter HQ algorithm using VGStudio (Volume Graphics, Germany). Scale bar: 10 μm. (**b**) The neuronal network was reconstructed in Cartesian coordinate space by tracing structures in the image. First, neurites were scanned by calculating the gradient vector flow (Xu et al., 1998) and then traced using a three-dimensional Sobel filter (Al-Kofahi et al., 2002). The resultant computer-generated model was examined and edited as reported previously (Mizutani et al., 2019). The obtained model of the S1A dataset was drawn using MCTrace (Mizutani et al., 2013). Structural constituents are color-coded. The same analysis was repeated for 34 datasets of 4 schizophrenia and 4 control cases. (**c**) Three-dimensional cage representation of the observed image (gray) is superposed on the structural model of an apical dendrite (red) indicated with a small dashed box in panel **b**. The three-dimensional map is contoured at 3.0 times the standard deviation

(3.0 σ) of the voxel values with a grid size of 97.6 nm. Nodes composing the structure are indicated with octagons. (**d**) Three-dimensional Cartesian coordinates of the traced structure were used to calculate geometric parameters, including average curvature and average torsion for each neurite. A volume indicated with the lower box in panel **b** is magnified and its two representative neurites are highlighted, of which the blue one is tortuous and hence has a high average curvature, while the black one is rather straight and hence has a low average curvature. Arrows indicate corresponding positions in the beeswarm plot that shows the curvature distribution of all neurites in this S1A structure.

The obtained results of the BA22 temporal cortex indicated that the structural parameters calculated from each dataset vary depending on the cases (Fig. 2a, 2b; significance was defined as $p < 0.05$). This suggests that the BA22 of each individual has its own cellular geometry. Next, we compared the structural parameters of the BA22 and BA24 cortexes and found that the frequency distribution of neurite curvature significantly differs between the brain areas for both the schizophrenia and control cases except for the schizophrenia S3 case (Fig. 2c). In contrast, the frequency distribution of neurite torsion showed similar zero-centered profiles for all cases (Supplementary Data S4). The magnitude relation of neurite curvatures of BA22 and BA24 is not consistent and varies from case to case. In the control N2 case for example, the mean neurite curvature of BA22 (0.58 um$^{-1}$; Supplementary Table S1 and S4) is 1.31 times higher than that of BA24 (0.44 um$^{-1}$), while in the control N4 case, the mean neurite curvature of BA22 (0.35 um$^{-1}$) is 0.84 times lower than that of BA24 (0.41 um$^{-1}$). The distribution profiles themselves vary depending on the brain area. In the N3 control case for example, the frequency distribution of the neurite curvature of BA22 shows a diamond-shape profile in contrast to the drop-shaped profile of BA24 (Fig. 2c), indicating that low-curvature neurites are rather minor in BA22 compared to BA24. The frequency distributions of the mean thickness radii and lengths of dendritic spines also indicate significant differences between BA22 and BA24 for both the schizophrenia and control cases, except for the control N1 case (Fig. 2d, Supplementary Data S4). The frequency distribution of the spine length of BA22 shows a bell-shaped profile in every case, while that of BA24 shows drop-shaped profiles in the S1 and the N3 cases or long-upper tails in the S4 and N2 cases (Supplementary Data S4). This results in a significant difference in mean spine length between the BA22 and the BA24 areas (Fig. 2e). These structural differences of neurites and spines between brain areas and between cases indicate that 1) neuron structures are dissimilar between the brain areas in every individual and that 2) the dissimilarity itself varies from person to person. The neurite curvature determines the spatial trajectory of neurites and hence affects neuronal microcircuits. The spine radius and length determine the local synaptic potential and hence affect the electrophysiological activity of neurons (Spruston, 2008). Therefore, the area/case dependency of neuron structures can influence the functional performance of their brain areas; e.g., a certain person may be good at thinking, while another person may be good at hearing.

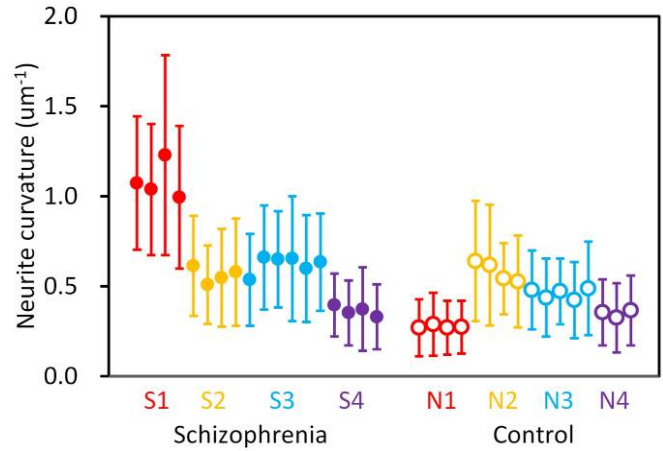

(**a**)

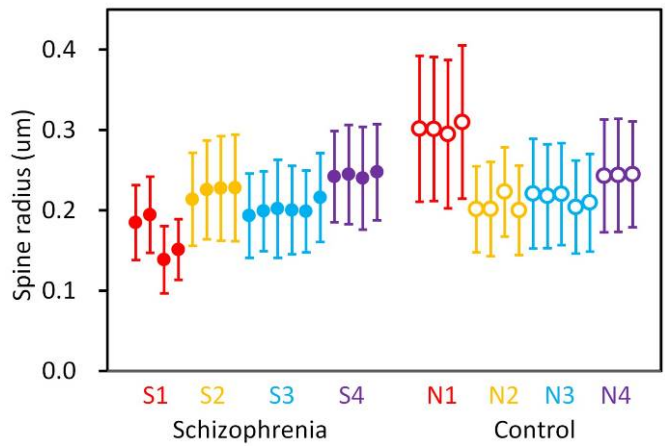

(**b**)

**Figure 2.** Differences in neuron structure between brain areas and between cases. (**a**,**b**) Neurite curvature (**a**) and spine thickness radius (**b**) of temporal (BA22) cortex calculated from each dataset vary depending on the cases ($p = 2.9 \times 10^{-8}$ for neurite curvature, $p = 4.5 \times 10^{-8}$ for spine radius, Welch's ANOVA with case as the main factor). Schizophrenia cases S1–S4 and age/gender-matched control cases N1–N4 are color-coded. Neurite curvature was calculated by averaging the curvature along the neurite. Spine radius was calculated by averaging the thickness radius along the dendritic spine. Circles indicate mean values of each dataset, and error bars indicate standard deviation.

(c)

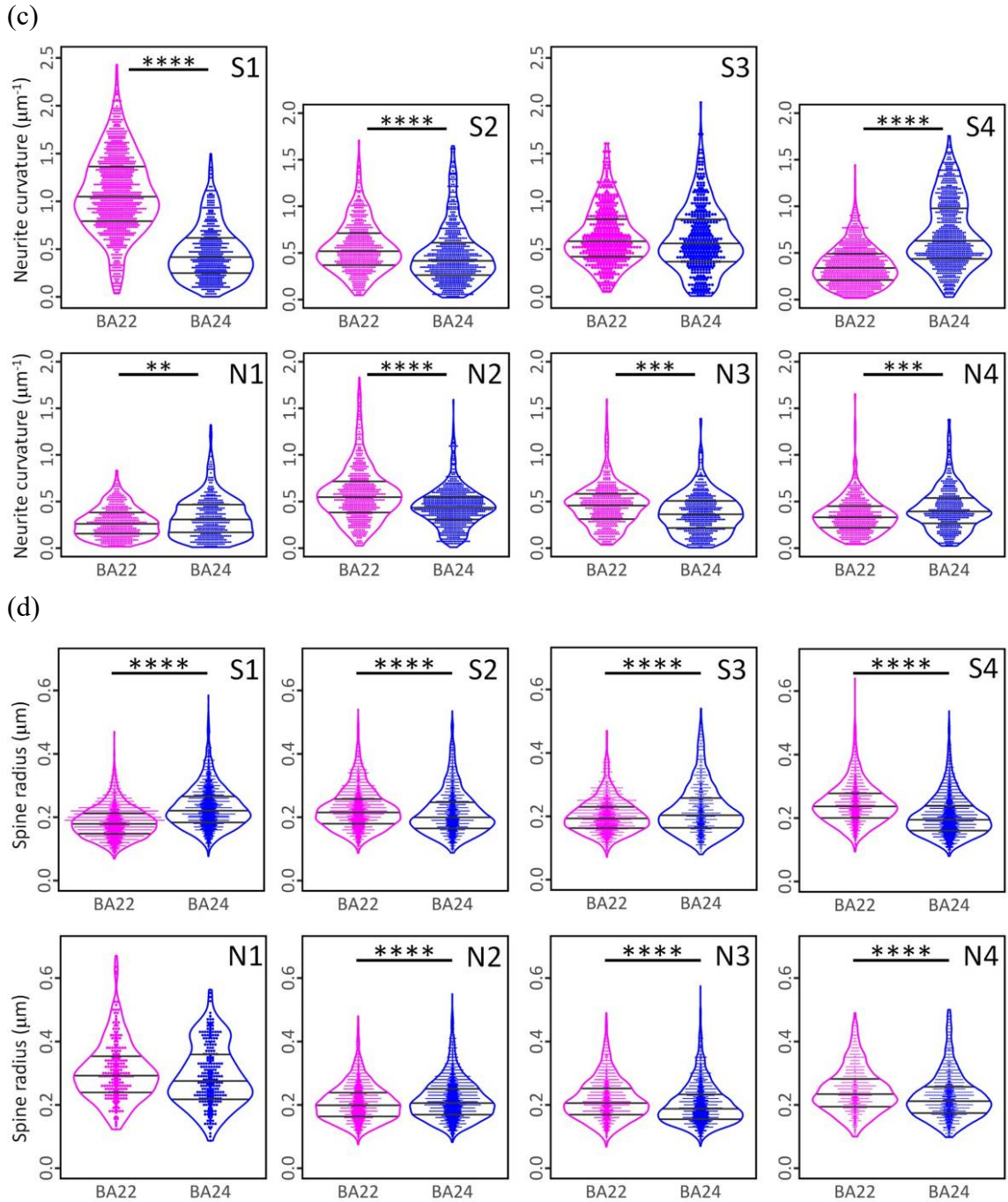

(d)

**Figure 2 (cont'd). (c,d)** Frequency distribution of neurite curvature (**c**) and spine radius (**d**) of temporal (BA22) and prefrontal (BA24) cortexes of each case. Schizophrenia cases S1–S4 and controls N1–N4 are indicated with labels. The equality of distributions between BA22 and BA24 was examined using the Kolmogorov-Smirnov test, and their *p*-values were corrected with the Holm-Bonferroni method. ****$p < 10^{-8}$; ***$p < 10^{-4}$; **$p < 10^{-2}$. Quartiles are indicated with bars. Dot size is adjusted for visibility.

The geometric analyses also revealed structural differences between schizophrenia cases and controls. The frequency distribution of neurite curvature of the schizophrenia cases shows fat upper tails (Fig. 2c), resulting in a significantly higher mean curvature in the schizophrenia cases than in the controls (Fig. 2f). Consequently, the neurites of the schizophrenia cases are tortuous and thin (Fig. 2g), while the neurites of the control cases are smooth and thick (Fig. 2h). Since the thinning of the neurites and spines can reduce the tissue volume, the volumetric changes in schizophrenia brains (Wright et al., 2000; Olabi et al., 2011; Haijma et al., 2013) should originate from these microscopic structural changes in neurons. The neurite curvature of the schizophrenia cases show larger deviations between cases and between areas compared with the controls (Fig. 2i and 2j). In the schizophrenia S1 case for example, the mean neurite curvature of BA22 (1.08 um$^{-1}$; Supplementary Table S1) is 2.3 times higher than that of BA24 (0.46 um$^{-1}$), while in the schizophrenia S4 case, the mean neurite curvature of BA22 (0.36 um$^{-1}$) is 0.51 times lower than that of BA24 (0.71 um$^{-1}$). Mean spine radius shows magnitude relations opposite to those of neurite curvature. In the S1 case for example, the spine radius is thinner in BA22 than in BA24, while neurite curvature is higher in BA22 than in BA24 (Fig. 2d).

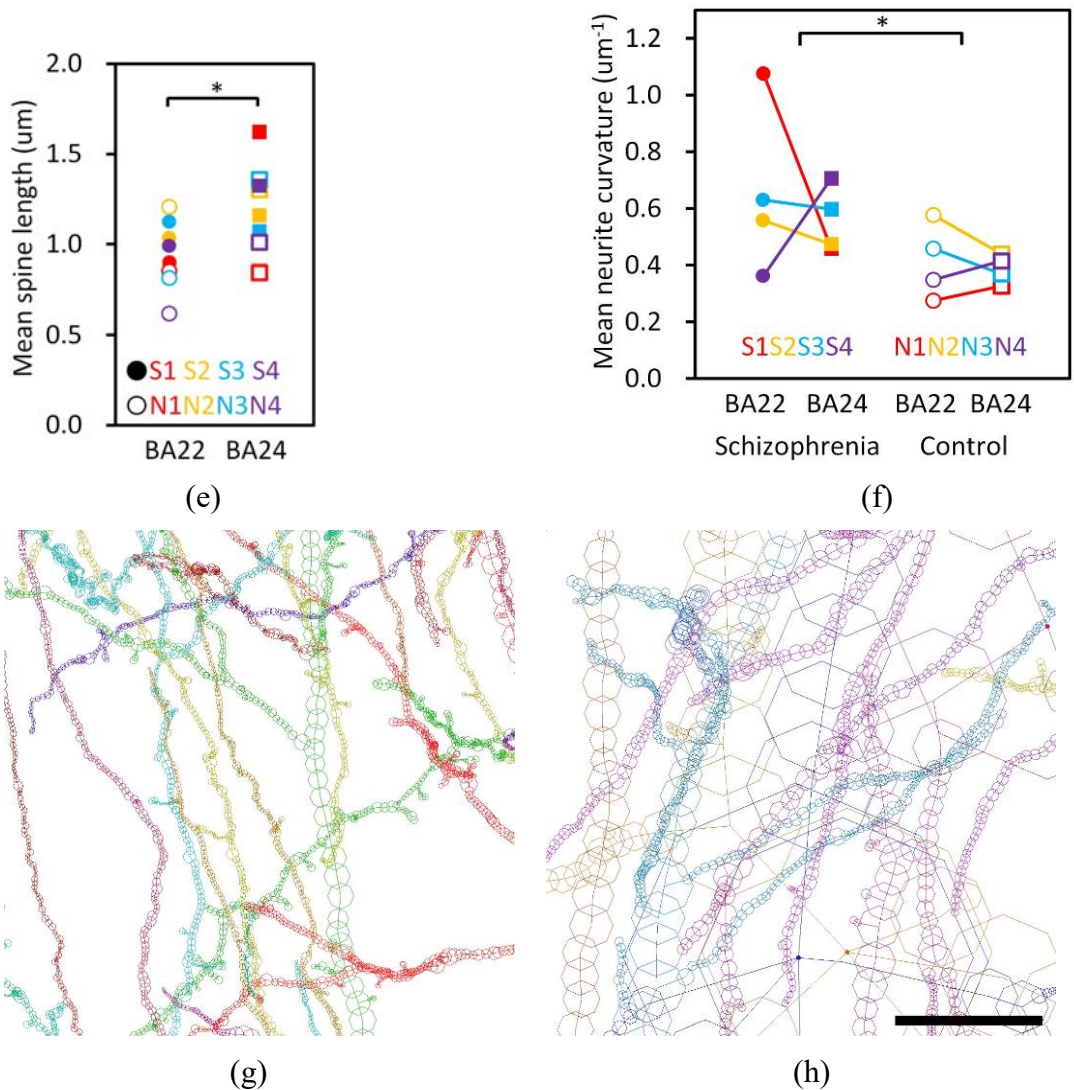

(e)

(f)

(g)

(h)

**Figure 2 (cont'd).** (**e**) Mean spine length is significantly shorter in BA22 than in BA24 (*$p$ = 0.023, two-way ANOVA with group (schizophrenia/control) and brain area (BA22/BA24) as main factors). Schizophrenia cases S1–S4 and controls N1–N4 are color-coded. (**f**) Mean neurite curvature is significantly higher in the schizophrenia cases than in the controls (*$p$ = 0.031, two-way ANOVA with group (schizophrenia/control) and brain area (BA22/BA24) as main factors). (**g**) Neurites of BA22 of the schizophrenia S1 case are tortuous and thin, (**h**) while those of the age/gender-matched control N1 case are smooth and thick. Structures of S1A and N1A are drawn to the same scale using MCTrace (Mizutani et al., 2013). The pial surface is toward the top. Structural constituents are color-coded. Nodes composing each constituent are indicated with octagons. The color coding and structural orientation are the same as in Fig 1b and Supplementary Data S3a. Scale bar: 10 μm.

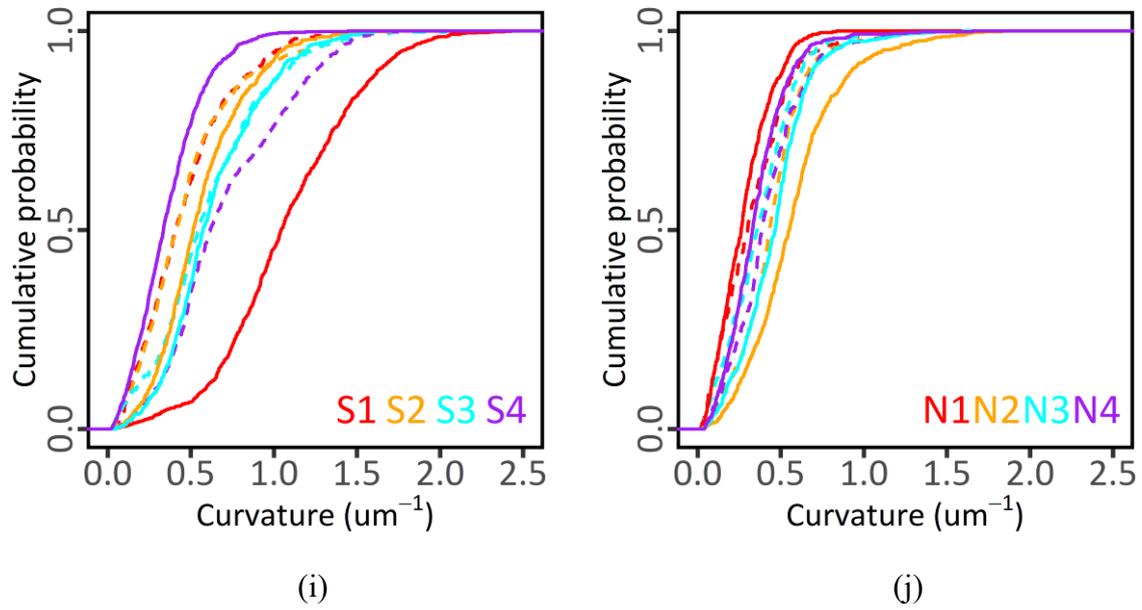

(i)

(j)

**Figure 2 (cont'd).** (**i,j**) Cumulative distribution of neurite curvature of schizophrenia cases (**i**) and controls (**j**). Solid lines represent BA22 distributions, and dashed lines represent BA24 distributions. Cases are color-coded.

Common structural features of neurons across brain areas and across cases were also identified from the structural analysis. Fig. 3a shows the relationship between mean thickness radius of neurites and that of spines. The plot shows a linear correlation, indicating that mean spine thickness correlates with mean neurite diameter. This suggests that these functionally different elements are structurally constrained in the neuron. Another common feature of the neuron structure was found in the relationship between curvature and thickness. Since a thin thread can sharply curve whereas a thick thread cannot, curvature is reciprocal to thickness radius as a general rule. A plot of this relationship (Fig. 3b) shows in all the cases that both neurites and spines follow the same reciprocal relationship independently of the brain area. This suggests that both neurites and spines follow a certain common physical or geometric principle that governs their structural constitution.

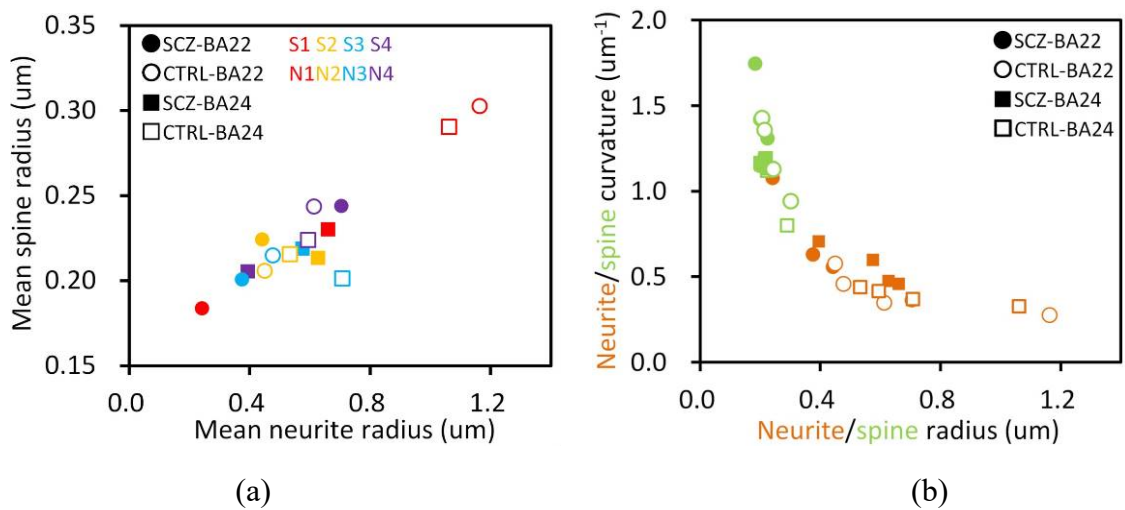

(a)  (b)

**Figure 3.** (**a**) Linear correlation between mean thickness radii of neurites and spines. Arithmetic mean of neurite radii or spine radii was calculated for BA22 or BA24 of each case. Schizophrenia cases S1–S4 and controls N1–N4 are color-coded. (**b**) Reciprocal relationship between curvature and thickness radius. Arithmetic mean of curvature or radius of neurites (orange) or spines (green) was calculated for BA22 or BA24 of each case.

This study revealed the structural diverseness of neurons between brain areas and between cases. The geometric analysis identified inter-area/case differences of neurite curvature (Fig. 2c,f) that reciprocally correlates with neurite thickness (Fig. 3b). According to the cable theory, neurite diameter affects the propagation distance of the action potential. Indeed, it has been reported that the neurite diameter is related with the action potential amplitude (Hansson et al., 1994). The linear correlation of neurite thickness to that of spines (Fig. 3a) indicates that neurite thinning narrows spines. Spine size associates with the long-term potentiation (Yang et al., 2008) and also with spine substructures, such as postsynaptic density (Arellano et al., 2007; Bosch et al., 2014). Therefore, thinning of neurites and spines alters the firing efficiency of neurons (Spruston, 2008) and affects the activity of their brain area. The geometric analysis also indicated that these functionally relevant structures differ from area to area and from case to case (Fig. 2). This coincides with the fact that each individual has a unique functional spectrum in their brain and hence has his or her own talents.

The schizophrenia cases showed a thin and tortuous neuronal network compared with the controls (Fig. 2f-h), suggesting that the neuron structure is associated with the disorder. The diverseness of neurite curvature between brain areas was rather large in the schizophrenia cases (Fig. 2i,j). The S1 case showed the highest neurite curvature in BA22 but a rather low curvature in BA24 (Fig. 2f). In contrast, the relationship is reversed in the S4 case: the neurite curvature of BA22 is low, while that of BA24 is high. These results imply that the brain areas compensate each other. However, the large heterogeneity of neurons in the schizophrenia cases can cause functional imbalances between brain areas that may result in disorders of total brain function. The large-scale heterogeneity of schizophrenia brains observed in MRI studies (Alnæs et al., 2019; Zhang et al., 2015) is ascribable to a microscopic heterogeneity of brain tissue. However, we have no consensus regarding the neuropathology of schizophrenia at present (Harrison, 1999). The area-dependent heterogeneous nature of the brain tissue of the schizophrenia cases leads to histological differences between cases and between brain areas and thereby gives rise to seemingly unreproducible results. We suggest that the controversy in the neuropathology of schizophrenia represents the area/case-dependent heterogeneity of neurons in schizophrenia.

The geometric diverseness of neurons found in this study cannot be delineated without conducting three-dimensional analyses of human brain tissues. An advantage of three-dimensional tissue analyses is that the structural difference can be quantitatively examined. Although neuron structures have been visualized using histological sections, structures along the viewing direction, such as spines behind neurites, are missed in light microscopy images, resulting in methodological biases in the subsequent analysis. In contrast, three-dimensional analyses allow us to examine exact structures by using mathematically defined parameters. A major limitation of this study is that the number of analyzed cases was limited by the availability of synchrotron radiation beamtime for the three-dimensional visualization. Although thousands of neurites and tens of thousands of spines were analyzed from their three-dimensional coordinates, the present results are based on brain tissues of just 4 schizophrenia and 4 control cases. The analysis should be repeated for additional cases in order to further examine the geometric diverseness of neurons. Another limitation of our analysis is the staining property of the Golgi method used for visualizing neurons in the x-ray image. Although the Golgi method has long been used for morphological studies of neurons, it does not stain every neuron and only visualizes a subset of neurons. Therefore, the analyzed neurons should be regarded as representative examples of the entire neuron population. It is also possible that antipsychotics caused the inter-area/case heterogeneity in the schizophrenia cases, though such an area-dependent heterogeneous action of therapeutics can be ascribed to 1) heterogeneous distribution of the therapeutics and/or 2) heterogeneous reaction of neurons to the therapeutics. Since these pharmacokinetic or physiological heterogeneities should be associated with the genetic mosaicism of human brain neurons (Lodato et al., 2015; McConnell et al., 2017), the relationship between the action of therapeutics and the genetic mosaicism should be further investigated in a number of schizophrenia cases.

Our nanometer-scale three-dimensional study of human brain tissues revealed the structural diverseness of neurons between brain areas and between cases. The brain has a modular architecture (Gómez-Robles et al., 2014; Bertolero et al., 2015) in which different areas with different genetic backgrounds (Lodato et al., 2015; McConnell et al., 2017) exert different functions (Amunts & Zilles, 2015). Besides the genetic mosaicism, environmental factors and education differences between individuals provide different stimuli to different brain areas, possibly resulting in differences in neuron structures between brain areas. The neurite curvature of the S4 case carrying a frameshift mutation in the *GLO1* gene was highest for BA24 (Mizutani et al., 2019), but lowest for BA22 among the schizophrenia cases (Fig. 2f). These results indicate that the neurite curvature is not ascribable solely to a single mutation. Rather, the structural diverseness between brain areas should have originated from multiple factors that affected neuron structures throughout the life of each individual. We suggest that further studies on neuron geometry and its relation with person-associated factors will shed light on the physical basis of the individuality of our brains.

**Methods**

**Cerebral tissue samples**

    All post-mortem human cerebral tissues were collected with informed consent from the legal next of kin using protocols approved by the Clinical Study Reviewing Board of Tokai University School of Medicine (application no. 07R-018) and the Ethics Committee of Tokyo Metropolitan Institute of Medical Science (approval no. 17-18). This study was conducted under the approval of the Ethics Committee for the Human Subject Study of Tokai University (approval nos. 11060, 11114, 12114, 13105, 14128, 15129, 16157, 18012, and 19001). The schizophrenia patients S1–S4 and control cases N1–N4 (Supplementary Table S1) of this study were the same with those analyzed in our previous report on the Brodmann area (BA) 24 (Mizutani et al., 2019). The number of cases was determined by the available beamtime at the synchrotron radiation facilities. Cerebral tissues of BA22 of the superior temporal gyrus were collected from the left hemispheres of the biopsied brains and subjected to Golgi impregnation (Mizutani et al., 2008a). Our previous results (Mizutani et al., 2008a; Mizutani et al., 2010; Mizutani et al., 2019) indicated that the staining procedure used in this study mainly visualizes neurons and blood vessels. The Golgi-stained tissues were then embedded in borosilicate glass capillaries using epoxy resin, as described previously (Mizutani et al., 2019).

**Microtomography and nanotomography**

    Overall tissue structures were visualized with simple projection microtomography at the BL20XU beamline (Suzuki et al., 2004) of SPring-8, using monochromatic radiation at 12 keV (Supplementary Table S2). Absorption contrast images of the brain tissues were recorded with a CMOS-based imaging detector (ORCA-Flash4.0, Hamamatsu Photonics, Japan). The layer V position (Supplementary Table S1) of the N3 tissue was estimated from the obtained microtomographic image. The layer V positions of other tissues were estimated from the microtomographic images along with Nissl sections. Examples of the microtomographic images and the Nissl sections are shown in Supplementary Data. S5. The data collection conditions are summarized in Supplementary Table S2.

    Nanotomography experiments using Fresnel zone plate optics were performed at the BL37XU beamline (Suzuki et al., 2016) of the SPring-8 synchrotron radiation facility and at the 32-ID beamline (De Andrade et al., 2016) of the Advanced Photon Source (APS) of Argonne

National Laboratory, as reported previously (Mizutani et al., 2019). Examples of raw image and reconstructed slice are shown in Supplementary Data S6. Photon flux at the sample position was measured using $Al_2O_3$:C dosimeters (Nagase-Landauer, Japan). Spatial resolutions were estimated using three-dimensional test patterns (Mizutani et al., 2008b) or from the Fourier domain plot (Mizutani et al., 2016). The experimental conditions are summarized in Supplementary Table S2.

**Structural analysis**

Tomographic slices were reconstructed with the convolution-back-projection method using the RecView software, as reported previously (Mizutani et al., 2019). The reconstruction calculation was performed by RS. The reconstructed datasets were then analyzed with the role allotment of data management to RS and data analysis to RM. RS provided the datasets to RM in 3 batches without the case information in order to eliminate human biases in the model building. RM analyzed the structure according to the method of our previous study (Mizutani et al., 2019) and returned the number of neurites of each dataset to RS without acknowledging the kind of structure represented with that number. No other results were disclosed to RS at this point. RS aggregated the number according to the case information in order to find cases in which aggregated numbers were less than 500. The aggregation results were not returned to RM. Datasets to be further analyzed were chosen by RS from predefined candidates and provided to RM without the case information.

After RM finished building models for all of the datasets, all coordinate files of the structural models in Protein Data Bank format were locked down. Then the case information was disclosed to RM. Two dummy datasets unrelated to this study were included in order to shuffle datasets taken at the 32-ID beamline. All datasets except these two dummy sets were used in the subsequent analysis. A total of 34 geometric data from the 4 schizophrenia and the 4 control cases were grouped according to the case information and used for the geometric analysis. We found that the data blinding was not essential because the obtained results were unpredictable from the case information.

In these procedures, Cartesian coordinate models were built using the MCTrace software (Mizutani et al., 2013). Geometric parameters were calculated from the obtained structures using the same software. Although user interfaces of MCTrace were updated from the version used in our previous study (Mizutani et al., 2019), we confirmed that the updated and the previous versions gave the same results. The statistics of the obtained structures are summarized in Supplementary Table S3.

## Code availability

RecView software (Mizutani et al., 2010) and its source code used for the tomographic reconstruction are available from https://mizutanilab.github.io under the BSD 2-Clause License. The model building and geometric analysis procedures were implemented in the MCTrace software available from the same site under the BSD 2-Clause License. Its source code is available upon request.

## Statistical tests

Statistical tests of structural parameters were performed using the R software. Significance was defined as $p < 0.05$. Differences in structural parameters between cases were examined using Welch's ANOVA. Mean neurite curvature and dendritic spine length of the BA22 and BA24 areas of schizophrenia and control cases were analyzed using two-way ANOVA with group (schizophrenia vs. control) and brain area (BA22 vs. BA24) as main factors. The equality of the probability distributions was examined using two-sided Kolmogorov-Smirnov tests, and the resultant $p$-values were corrected with the Holm-Bonferroni method.

## Conflict of interest

MI and MA declare a conflict of interest, being authors of several patents regarding therapeutic use of pyridoxamine for schizophrenia. All other authors declare no conflict of interest.

## Acknowledgements

We are grateful to Prof. Motoki Osawa and Akio Tsuboi (Tokai University School of Medicine) for their generous support of this study. We thank Noboru Kawabe (Support Center for Medical Research and Education, Tokai University) for assistance in preparing the histology sections. We also thank the Technical Service Coordination Office of Tokai University for assistance in preparing adapters for nanotomography. This work was supported by Grants-in-Aid for Scientific Research from the Japan Society for the Promotion of Science (nos. 21611009, 25282250, and 25610126), and by the Japan Agency for Medical Research and Development under grant nos. JP18dm0107087, JP19dm0107087, JP18dm0107088, JP19dm0107088h0004, JP20dm0107088h0005, and JP19dm0107108. The analysis of human tissues was conducted under the approval of the Ethics Committee for the Human Subject Study of Tokai University, the Clinical Study Reviewing Board of Tokai University School of Medicine, the Ethics Committee of Tokyo Metropolitan Institute of Medical Science, and the Institutional Biosafety Committee of Argonne National Laboratory. The synchrotron radiation experiments at SPring-8 were performed with the approval of the Japan Synchrotron Radiation Research Institute (JASRI) (proposal nos. 2011A0034, 2015A1160, 2015B1101, 2016B1041, 2018A1164, and

**Author Contributions**

RM designed the study. RM, RS, ST, CI, NN, YT, IK, SI, NO, KO, MI, and MA prepared the samples. RM, RS, MU, AT, KU, YT, YS, VDE, and FDC performed the synchrotron radiation experiments. RM, RS, and YY analyzed the data. RM wrote the manuscript and prepared the figures. All authors reviewed the manuscript.

**References**

1. Brodmann K. Vergleichende Lokalisationslehre der Großhirnrinde in ihren Prinzipien dargestellt auf Grund des Zellenbaues, Leipzig: Johann Ambrosius Barth, 1909.

2. Amunts K, Zilles K. Architectonic mapping of the human brain beyond Brodmann. *Neuron* 2015; **88**: 1086–1107.

3. Sporns O, Betzel RF. Modular Brain Networks. *Annu Rev Psychol.* 2016; **67**: 613–640.

4. Mars RB, Passingham RE, Jbabdi S. Connectivity fingerprints: From areal descriptions to abstract spaces. *Trends Cogn Sci* 2018; **22**: 1026-1037.

5. Finn ES *et al.* Functional connectome fingerprinting: identifying individuals using patterns of brain connectivity. *Nat Neurosci* 2015; **18**: 1664–1671.

6. Passingham RE, Stephan KE, Kötter R. The anatomical basis of functional localization in the cortex. *Nat Rev Neurosci* 2002; **3**: 606–616.

7. Lodato MA, Woodworth MB, Lee S, Evrony GD, Mehta BK, Karger A, Lee S, Chittenden TW, D'Gama AM, Cai X, Luquette LJ, Lee E, Park PJ, Walsh CA. Somatic mutation in single human neurons tracks developmental and transcriptional history. *Science* 2015; **350**: 94–98.

8. McConnell MJ, Moran JV, Abyzov A, Akbarian S, Bae T, Cortes-Ciriano I, Erwin JA, Fasching L, Flasch DA, Freed D, Ganz J, Jaffe AE, Kwan KY, Kwon M, Lodato MA, Mills RE, Paquola ACM, Rodin RE, Rosenbluh C, Sestan N, Sherman MA, Shin JH, Song S, Straub RE, Thorpe J, Weinberger DR, Urban AE, Zhou B, Gage FH, Lehner T, Senthil G, Walsh CA, Chess A, Courchesne E, Gleeson JG, Kidd JM, Park PJ, Pevsner J, Vaccarino FM; Brain Somatic Mosaicism Network. Intersection of diverse neuronal genomes and neuropsychiatric disease: The Brain Somatic Mosaicism Network. *Science* 2017; **356**: eaal1641.

9.  Mizutani R, Saiga R, Takeuchi A, Uesugi K, Terada Y, Suzuki Y, De Andrade V, De Carlo F, Takekoshi S, Inomoto C, Nakamura N, Kushima I, Iritani S, Ozaki N, Ide S, Ikeda K, Oshima K, Itokawa M, Arai M. Three-dimensional alteration of neurites in schizophrenia. *Transl Psychiatry* 2019; **9**: 85.

10. Botvinick M, Nystrom LE, Fissell K, Carter CS, Cohen JD. Conflict monitoring versus selection-for-action in anterior cingulate cortex. *Nature* 1999; **402**: 179–181.

11. Bush G, Luu P, Posner MI. Cognitive and emotional influences in anterior cingulate cortex. *Trends Cogn Sci* 2000; **4**: 215–222.

12. Bouras C, Kövari E, Hof PR, Riederer BM, Giannakopoulos P. Anterior cingulate cortex pathology in schizophrenia and bipolar disorder. *Acta Neuropathol* 2001; **102**: 373–379.

13. Fornito A, Yücel M, Dean B, Wood SJ, Pantelis C. Anatomical abnormalities of the anterior cingulate cortex in schizophrenia: bridging the gap between neuroimaging and neuropathology. *Schizophr Bull* 2009; **35**: 973–993.

14. Suzuki Y, Takeuchi A, Terada Y, Uesugi K, Mizutani R. Recent progress of hard x-ray imaging microscopy and microtomography at BL37XU of SPring-8. *AIP Conference Proceedings* 2016; **1696**: 020013.

15. De Andrade V *et al.* Nanoscale 3D imaging at the Advanced Photon Source. *SPIE Newsroom* 2016; doi: 10.1117/2.1201604.006461.

16. Takeuchi A, Uesugi K, Takano H, Suzuki, Y. Submicrometer-resolution three-dimensional imaging with hard x-ray imaging microtomography. *Rev Sci Instrum* 2002; **73**: 4246–4249.

17. Spruston N. Pyramidal neurons: dendritic structure and synaptic integration. *Nat Rev Neurosci* 2008; **9**: 206–221.

18. Wright IC *et al.* Meta-analysis of regional brain volumes in schizophrenia. *Am J Psychiatry* 2000; **157**: 16–25.

19. Olabi B *et al.* Are there progressive brain changes in schizophrenia? A meta-analysis of structural magnetic resonance imaging studies. *Biol Psychiatry* 2011; **70**: 88–96.

20. Haijma SV *et al.* Brain volumes in schizophrenia: a meta-analysis in over 18,000 subjects. *Schizophr Bull* 2013; **39**: 1129–1138.

21. Hansson BS, Hallberg E, Löfstedt C, Steinbrecht RA. Correlation between dendrite diameter and action potential amplitude in sex pheromone specific receptor neurons in male *Ostrinia nubilalis* (Lepidoptera: Pyralidae). *Tissue Cell* 1994; **26**: 503–512.

22. Yang Y, Wang XB, Frerking M, Zhou Q. Spine expansion and stabilization associated with long-term potentiation. *J Neurosci* 2008; **28**: 5740–5751.

23. Arellano JI, Benavides-Piccione R, Defelipe J, Yuste R. Ultrastructure of dendritic spines: correlation between synaptic and spine morphologies. *Front Neurosci* 2007; **1**: 131–143.

24. Bosch M, Castro J, Saneyoshi T, Matsuno H, Sur M, Hayashi Y. Structural and molecular

remodeling of dendritic spine substructures during long-term potentiation. *Neuron* 2014; **82**: 444–459.

25. Alnæs D, Kaufmann T, van der Meer D, Córdova-Palomera A, Rokicki J, Moberget T, Bettella F, Agartz I, Barch DM, Bertolino A, Brandt CL, Cervenka S, Djurovic S, Doan NT, Eisenacher S, Fatouros-Bergman H, Flyckt L, Di Giorgio A, Haatveit B, Jönsson EG, Kirsch P, Lund MJ, Meyer-Lindenberg A, Pergola G, Schwarz E, Smeland OB, Quarto T, Zink M, Andreassen OA, Westlye LT; Karolinska Schizophrenia Project Consortium. Brain Heterogeneity in Schizophrenia and Its Association With Polygenic Risk. *JAMA Psychiatry* 2019; **76**: 739–748.

26. Zhang T, Koutsouleris N, Meisenzahl E, Davatzikos C. Heterogeneity of structural brain changes in subtypes of schizophrenia revealed using magnetic resonance imaging pattern analysis. *Schizophr Bull* 2015; **41**: 74–84.

27. Harrison PJ. The neuropathology of schizophrenia: A critical review of the data and their interpretation. *Brain* 1999; **122**: 593–624.

28. Gómez-Robles A, Hopkins WD, Sherwood CC. Modular structure facilitates mosaic evolution of the brain in chimpanzees and humans. *Nat Commun.* 2014; **5**: 4469.

29. Bertolero MA, Yeo BT, D'Esposito M. The modular and integrative functional architecture of the human brain. *Proc Natl Acad Sci U S A* 2015; **112**: E6798–6807.

**References in figure caption**

30. Xu C, Prince JL. Snakes, shapes, and gradient vector flow. *IEEE Trans Image Process* 1998; **7**: 359–369.

31. Al-Kofahi KA *et al.* Rapid automated three-dimensional tracing of neurons from confocal image stacks. *IEEE Trans Inf Technol Biomed* 2002; **6**: 171–187.

32. Mizutani R, Saiga R, Takeuchi A, Uesugi K, Suzuki Y. Three-dimensional network of Drosophila brain hemisphere. *J Struct Biol* 2013; **184**: 271–279.

**References in the Methods section**

33. Mizutani R *et al.* Three-dimensional microtomographic imaging of human brain cortex. *Brain Res* 2008a; **1199**: 53–61.

34. Mizutani R *et al.* Microtomographic analysis of neuronal circuits of human brain. *Cereb Cortex* 2010; **20**: 1739–1748.

35. Suzuki Y *et al.* Construction and commissioning of a 248 m-long beamline with X-ray undulator light source. *AIP Conference Proceedings* 2004; **705**: 344–347.

36. Mizutani R, Takeuchi A, Uesugi K, Suzuki Y. Evaluation of the improved three-dimensional resolution of a synchrotron radiation computed tomograph using a micro-fabricated test pattern. *J Synchrotron Radiat* 2008b; **15**: 648–654.

37. Mizutani R *et al.* A method for estimating spatial resolution of real image in the Fourier domain. *J Microsc* 2016; **261**: 57–66.